# Molecular Diffusivity of Click Reaction Components: The Diffusion Enhancement Question

Nasrollah Rezaei-Ghaleh[†,# *], Jaime Agudo-Canalejo[‡], Christian Griesinger[†], Ramin Golestanian[‡*]

[†]Department of NMR-based Structural Biology, Max Planck Institute for Biophysical Chemistry, Am Faßberg 11, D-37077 Göttingen, Germany.

[#]Institut für Physikalische Biologie, Heinrich-Heine-Universität Düsseldorf, Universitätsstr. 1, D-40225 Düsseldorf, Germany.

[‡]Department of Living Matter Physics, Max Planck Institute for Dynamics and Self-Organization, Am Faßberg 11, D-37077 Göttingen, Germany.

**ABSTRACT:** Micrometer-sized objects are widely known to exhibit chemically-driven motility in systems away from equilibrium. Experimental observation of reaction-induced motility or enhancement in diffusivity at the much shorter length scale of small molecules is however still a matter of debate. Here, we investigate the molecular diffusivity of reactants, catalyst and product of a model reaction, the copper-catalyzed azide-alkyne cycloaddition click reaction, and develop new NMR diffusion approaches that allow the probing of reaction-induced diffusion enhancement in nano-sized molecular systems with higher precision than the state of the art. Following two different approaches that enable the accounting of time-dependent concentration changes during NMR experiments, we closely monitored the diffusion coefficient of reaction components during the reaction. The reaction components showed distinct changes in the diffusivity: while the two reactants underwent a time-dependent decrease in their diffusivity, the diffusion coefficient of the product gradually increased and the catalyst showed only slight diffusion enhancement within the range expected for reaction-induced sample heating. The decrease in diffusion coefficient of the alkyne, one of the two reactants of click reaction, was not reproduced during its copper coordination when the second reactant, azide, was absent. Our results do not support the catalysis-induced diffusion enhancement of the components of click reaction and, instead, point to the role of a relatively large intermediate species within the reaction cycle with diffusivity lower than both the reactants and product molecule.

Molecular machines which convert chemical energy into kinetic energy or mechanical work are key players in natural and synthetic biology and nanotechnology.[1-3] Of particular interest to applications such as drug delivery and nanorobotics is the transduction of chemical energy into translational motion in the bulk of a fluid.[4, 5] Artificial microscopic particles such as bimetallic rods, Janus particles and enzyme-coated beads are known to undergo self-propelled directed motion powered by their surface catalytic activity, with mechanisms that are by now well understood.[6, 7] Over long time scales, this ballistic motion is randomized by rotational diffusion, leading to greatly enhanced diffusive behavior.[8, 9] There have also been theoretical proposals for achieving stochastic swimming at the nano-scale by breaking the detailed balance akin to how biological molecular motors function.[10, 11] More recently, experiments have reported that also single enzymes may experience catalysis-induced enhanced diffusion,[12-14] although the possible underlying mechanisms and even the existence of this phenomenon are still under debate.[15-19] Continuing the quest towards translational motion at increasingly smaller scales, it was later claimed that even molecular-scale systems (a Grubbs catalyst) exhibit enhanced diffusion during catalysis,[20, 21] although this was subsequently shown to be due to a convection artifact in the measurements.[22]

A recent report, however, has reinvigorated the idea of enhanced diffusion during molecular catalysis.[23] In these experiments, it is claimed that the mobility of reactant molecules in a family of organic chemical reactions, including the Cu(I)-catalyzed Azide-Alkyne Cycloaddition (CuAAC) reaction, are boosted during catalysis.[23] With the word 'boosted', it is implied that the underlying mechanism is an active, propulsive one, akin to stochastic swimming[10, 11] associated with the (free) energy released during each catalytic event. From a theoretical perspective, however, propulsive motion can only be observed over a very short period of time, as the negligible inertial effects will lead to rapid dissipation of the kinetic energy from sudden 'kicks' into the environment, leading to a randomization of the propulsion by rapid rotational diffusion.[15, 17-19, 24] In addition to theoretical concerns, this report has been hotly debated from a technical perspective, especially with regards to challenges of diffusion measurement by NMR.[25-31]

Pulse Field Gradient (PFG) NMR is the only technique that can enable the monitoring of molecular diffusivity at atomic scale. The technique allows us to determine the diffusion coefficients for the various molecular components of a reaction mixture, including reactants, catalysts, intermediate species, products and solvent molecules. The PFG-NMR technique relies on spatial encoding of molecules via application of a magnetic field gradient along the z-axis of an NMR tube, through which the frequency of nuclear spins sitting on the molecules would carry z-coordinate information.[32] After a diffusion delay, during which the molecules undergo diffusion in different directions, including the z-axis, the spatial information is decoded through application of another field gradient pulse with the same magnitude but the opposite sign. The NMR signals of non-diffusing molecules will therefore be completely recovered after the second gradient pulse and its consequent reversing of nuclear spin frequencies, while the diffusing molecules will undergo NMR signal attenuation in dependence of their displacement along z-axis and the strength of magnetic field gradient. The NMR signal intensity *versus* gradient field strength data will then allow determining diffusion coefficients separately for each NMR-resolved signal and its underlying molecular species. Furthermore, when a chemical reaction takes place in the NMR sample, the real-time PFG-NMR experiments enables monitoring of molecular diffusion over the course of reaction. However, in such cases, proper technical adjustments in NMR diffusion experiments and data analysis should be made to account for NMR signal intensity changes due to the kinetics of reaction and consequent changes in the concentration, NMR relaxation properties etc.[33]

Here, we investigate molecular diffusion during the CuAAC reaction using the adjusted NMR diffusion experiments and introduce two ways in which artifacts due to time-dependent signal intensities in diffusion-NMR experiments can be identified and corrected, hence enabling detection and quantification of potential reaction-induced diffusion enhancements in nano-sized molecular systems. It is demonstrated that the two reactants, catalyst and product of this reaction experience distinct reaction-dependent alterations in their diffusivity. Our results do not support the uniform catalysis-induced diffusion enhancement in CuAAC reaction, as suggested in ref. [23], but instead, point to the role of intermediate species in the CuAAC reaction cycle as primary cause in reaction-dependent molecular mobility alterations.

The CuAAC reaction is one of the primary examples of click reactions which transforms organic azides and terminal alkynes into the corresponding 1,2,3-triazoles (scheme 1a).[34, 35] Since Cu(I) is the least thermodynamically stable oxidation state of copper, a combination of Cu(II) salt and ascorbate, a mild reducing agent, is often used as a source of Cu(I) in CuAAC reactions performed in aqueous solutions.[35] Unlike the uncatalyzed reaction which requires much higher temperatures and produces mixtures of 1,4- and 1,5-disubstituted triazole regioisomers, the copper-catalyzed reaction is fast at room temperature and produces nearly pure 1,4-disubstituted triazoles. This is because the catalysis by copper converts the mechanism of cycloaddition reaction into a sequence of discrete steps, where the activation energy barrier for the key rate-determining C-N bond formation is reduced compared to the uncatalyzed reaction.[36, 37] Much remains to be understood about the complex mechanism of the CuAAC reaction, but several lines of experimental evidence and DFT calculations propose that the reaction begins with the recruitment of a π-bound copper ion to the alkyne molecule, which acidifies its terminal proton and therefore facilitates its replacement with a second copper ion, hence formation of a σ-bond copper acetylide (scheme 1b, steps **I-II**). Then, the reversible coordination of the dinuclear copper intermediate with the azide molecule leads to the synergistic nucleophilic activation of the alkyne and electrophilic activation of the azide and drives the formation of the first C-N bond within a strained copper metallacycle (scheme 1b, steps **III-IV**). The subsequent energetically favorable steps of copper triazolide formation (step **V**) and copper substitution by proton (step **VI**) will then culminate in the formation of the triazole molecule as the reaction product.[36-41] Notably, the unique catalytic activity of copper ion is rooted in its combined propensity of engaging in π- and σ-interactions with terminal alkynes *and* rapid exchange of these and other ligand molecules (including solvent molecules), especially in aqueous solutions, in its coordination sphere.[36]

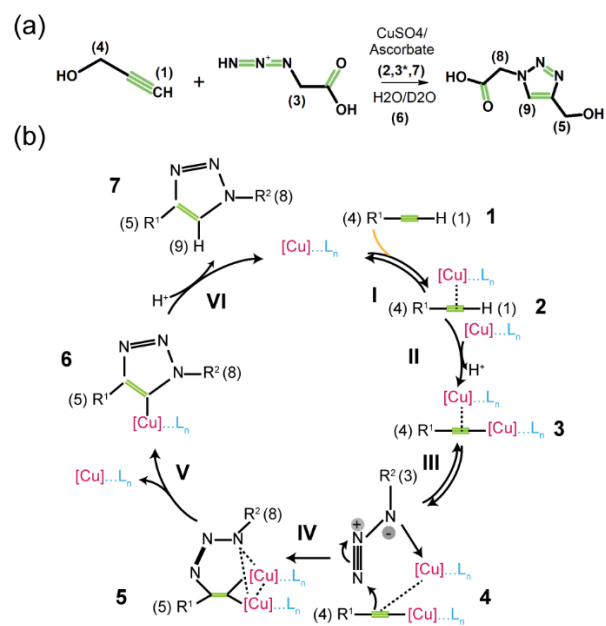

**Scheme 1.** (a) Copper-catalyzed azide-alkyne cycloaddition click reaction, with numbers (1)-(9) corresponding to the NMR signals studied here, and (b) schematic depiction of the catalytic cycle, proceeding through steps **I-VI**, involving chemical species **1-7**.

## RESULTS AND DISCUSSION

The 1D $^1$H NMR spectrum of the sample containing 0.2 M propyl-2-ny-ol (henceforth, alkyne or reactant 1) in D$_2$O is shown in Figure 1a (bottom panel). The two resonances of the alkyne molecule were observed at 4.238 and 2.838 ppm, respectively corresponding to the two methylene (-CH$_2$, signal #4) and one terminal (≡CH, signal #1) protons. Then, the diffusion coefficient of the alkyne molecule was measured via the PFG-NMR method. In general, the excellent linearity of the NMR signal intensity *vs* gradient field strength in log-quadratic scale, as illustrated in the Insets of Figure 1 (and SI, Figure S1), allowed precise determination of the diffusion coefficient of alkyne (and other) molecules. In addition, the convection-compensated PFG-NMR experiments confirmed that the contribution of convection to molecular mobility was negligible (see SI, Experimental Section).[42] The NMR diffusion measurement yielded the same diffusion coefficient ($D_0$, the subscript 0 denotes diffusion coefficient at equilibrium, i.e. in the absence of chemical reaction) of $11.1*10^{-10}$ m$^2$·s$^{-1}$ for signals #1 and #4, as expected for them sitting on the same diffusing molecule.

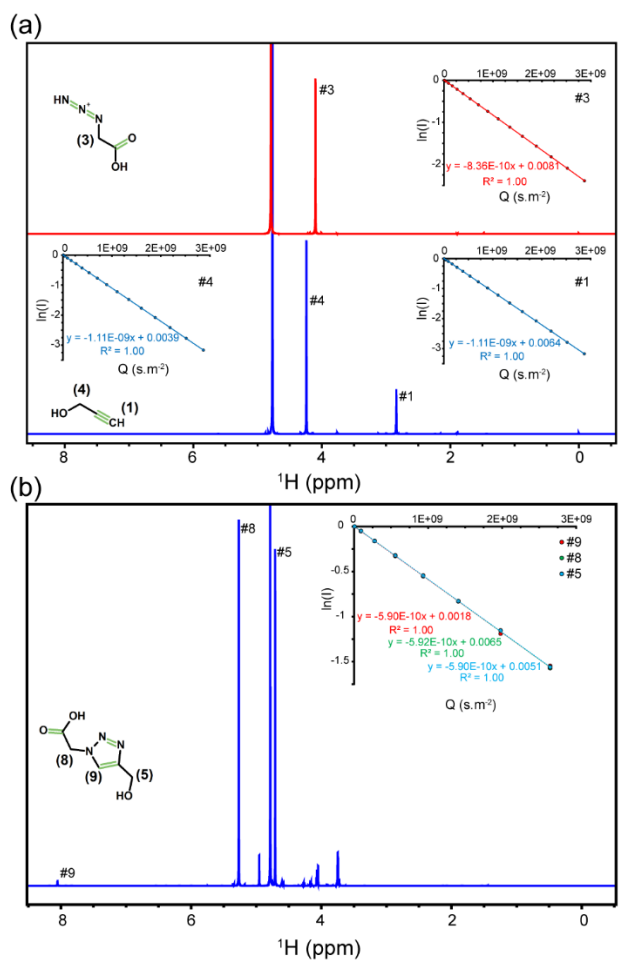

**Figure 1.** 1D $^1$H NMR spectra of the two reactants (a) and single product (b) of the click reaction measured in isolation (for reactants) or after completion of reaction (for product). The NMR signals are assigned according to the shown 2D chemical structures. The gradient-dependent NMR intensity attenuation of different NMR signals in log-quadratic scale are shown as Insets, in which the linear slopes represent diffusion coefficients of reactants and product molecules ($D_0$) used as reference in this study.

Figure 1a, top panel, shows the 1D $^1$H NMR spectrum of the sample containing 0.2 M 2-azido acetic acid (henceforth, azide or reactant 2) in D$_2$O. The resonance corresponding to the two methylene (-CH$_2$, signal #3) protons of the azide molecule was observed at 4.097 ppm and the associated $D_0$ was 8.4*10$^{-10}$ m$^2$.s$^{-1}$. The absolute $D_0$ of alkyne and azide molecules obtained here were slightly larger than the previously reported values (ca. 7 and 9%, respectively),[28] which is probably caused by small differences in temperature and/or imperfect compensation for convection artifacts, however the ratio between $D_0$ of alkyne and azide molecules was 1.32, in close agreement with the previously obtained value of 1.35.[28] The sample containing alkyne and azide molecules each at 0.2 M concentration in the absence of any catalyst did not show any considerable change in the chemical shifts of the three signals belonging to the two reactants or their associated diffusion coefficients. In addition, the NMR spectrum of the mixed sample did not change during an overnight incubation at 298 K, indicating that the rate of uncatalyzed click reaction was negligible at this temperature.

The 1D $^1$H NMR spectrum of the sample containing 64 mM sodium ascorbate (henceforth, ascorbate or catalyst) in D$_2$O is shown in SI, Figure S1a, where three signals at 4.529, 4.036 and 3.762 ppm, respectively corresponding to the methine (-CH) proton of the ring (signal #7) and the side chain (signal #3*), and the two methylene (-CH$_2$, signal #2) protons of the side chain of ascorbate molecules, were observed. The hydroxyl (-OH) protons are not expected to appear as separate resonances, as they are in rapid exchange with the solvent at this pH. The same $D_0$ of 5.8*10$^{-10}$ m$^2$.s$^{-1}$ was observed for all three signals (Fig. S1a, Inset). Addition of 16 mM CuSO$_4$ led to a slight displacement of peaks #2 and #3* towards downfield and alterations in their linewidth and splitting pattern, while peak #7 disappeared and four new peaks at chemical shifts of 4.707, 4.623, 4.308 and 4.202 ppm emerged (SI, Figure S1b). The peak at 4.707 ppm underwent gradual downfield displacement, indicating a chemical reaction occurring in the timescale of minutes (Fig. S1b, Inset). The newly emerged peaks belonged to the oxidation products of ascorbate, such as dehydroascorbic acid, induced in the presence of Cu(II) ions. After completion of reaction, the signals #2 and #3* exhibited the same diffusion coefficient of 5.9*10$^{-10}$ m$^2$.s$^{-1}$, very close to the $D_0$ value obtained for the ascorbate molecule before addition of CuSO$_4$. The newly emerged signals however showed slightly larger diffusion coefficients in the range of 6.0-6.4*10$^{-10}$ m$^2$.s$^{-1}$, consistent with them belonging to different molecular species generated through oxidation of ascorbate to smaller molecules.

Next, to obtain the diffusion coefficient of the 1,4-disubstituted 1,2,3-triazole molecule produced by the CuAAC reaction (henceforth, triazole or product), we started the reaction by adding 16 mM CuSO$_4$ and 64 mM sodium ascorbate to the mixture of alkyne and azide molecules, each at 200 mM concentration, and let the catalyzed reaction be completed during overnight incubation at 298 K. The 1D $^1$H NMR spectrum of the sample after reaction completion shows three signals at 8.050 (signal #9), 5.268 (signal #8) and 4.709 ppm (signal #5) belonging to the product triazole molecule (Figure 1b). The diffusion coefficients associated to the three product signals were similarly 5.9*10$^{-10}$ m$^2$.s$^{-1}$. Overall, the diffusion coefficients of the four molecules studied here followed the order $D_{alkyne}$(1.1*10$^{-9}$)>$D_{azide}$(0.84*10$^{-9}$)>$D_{triazole}$(0.59*10$^{-9}$)~$D_{ascorbate}$(0.58*10$^{-9}$), in qualitative agreement with the molecular-mass-based estimation of their diffusion coefficients (their molecular masses are 56.06, 101.06, 157.13 and 176.12 Da, respectively). Next, we monitored through real-time 1D $^1$H NMR experiments how the three proton signals of the alkyne and azide reactant molecules varied over the course of the CuAAC reaction triggered by addition of CuSO$_4$ and sodium ascorbate. As shown in Figure 2a, the signal #1 belonging to the terminal proton of the alkyne molecule underwent a time-dependent chemical shift displacement towards downfield along with a drastic decrease in signal intensity. The signals #4 and #3, belonging to the methylene protons of the alkyne and azide molecules respectively, experienced similar albeit smaller downfield chemical shift displacement, as well as intensity loss (Figures 2b-d). On the other hand, the signals #5, #8 and #9 belonging to the product molecule showed time-dependent intensity gain, which in case of signals #5 and #9 was accompanied with upfield chemical shift displacements, while a downfield chemical shift displacement was observed for signal #8 (Figures 3a-d). Interestingly, the upfield chemical shift displacement of signal #9 was partially reversed before the completion of reaction, suggesting the presence of multiple, i.e. more than two, chemical species underlying this signal. In general, the presence of single NMR signals per proton species of the reactant and product molecules indicates that the related exchange processes along the reversible steps of the reaction cycle are fast with respect to the relevant NMR chemical shift timescales. In addition, the NMR evidence for the presence of multiple species underlying signal #9 indicates that the reaction mechanism is more complex than what is shown in scheme 1b.

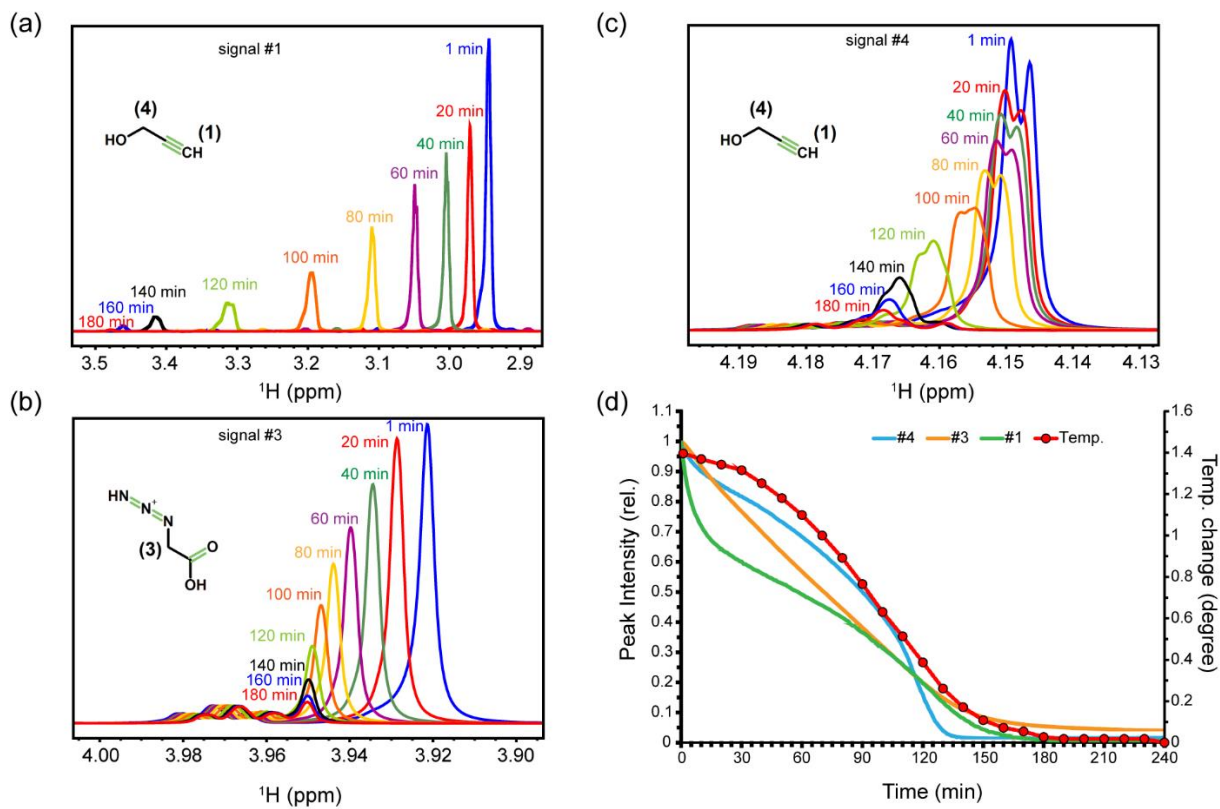

**Figure 2.** Kinetics of reactant consumption during click reaction monitored through real-time 1D $^1$H NMR spectra. Time-dependent changes in signals #1 (terminal proton of alkyne), #3 (methylene protons of azide) and #4 (methylene protons of alkyne) are shown in panels (a)-(c), along with 2D chemical structures of the reactant molecules. Time-dependent changes in NMR signal intensities are shown in (d). Note the difference in the kinetic profiles of the three signals, reflecting their different entry points to the click reaction catalytic cycle. In (d), time-dependent changes in the temperature of the NMR sample during click reaction are shown with respect to the equilibrium temperature.

Subsequently, we monitored how the temperature of the NMR sample changes along the CuAAC reaction. As evaluated through the chemical shift difference between the reference tetramethylsilane (TMS) and residual water (HDO) proton signals and its temperature dependence,[43] the (average) temperature of the NMR sample was higher by around 1.4 °C in the beginning of the click reaction and gradually decreased during the first steps of reaction cycle, as shown previously,[36-38] and implies that the heat generated within the NMR tube could not be dissipated instantaneously. Based on the temperature dependence of water viscosity, the average rise of 1.4 °C in the temperature reduces the sample viscosity by 3.1% and is therefore expected to increase the diffusion coefficients by ca. 3.6%.

Next, we employed real-time PFG-NMR diffusion measurements in order to monitor how the CuAAC reaction influences the molecular mobility of reactants, product and catalyst molecules. Since the NMR signals of reactants and product molecules underwent considerable intensity changes during NMR diffusion measurements, as illustrated in Figures 2 and 3, we needed to account for the reaction time-dependent signal intensity changes in addition to the gradient field-dependent intensity changes. Otherwise, it would lead to a systematic over- or under-estimation of diffusion coefficients derived through the standard Stejskal-Tanner (ST) equation (SI, eq. S1), depending on the decreasing or increasing trend of signal intensities, respectively.[33] To avoid such artifacts, we employed an approach in which, unlike the standard PFG-NMR diffusion experiment in which the gradients are ordered in increasing strength, the order of gradient strengths was shuffled in a way that the correlation between reaction time (and its consequent signal intensity changes) and gradient strength approached zero. This approach eliminates the systematic error in diffusion coefficients due to kinetic effects, although it may enhance the scattering of intensity *vs* gradient field strength data and therefore increase the random error. To reduce the potential random error caused by gradient shuffling, we utilized the scan-interleaved NMR experimental scheme so that the time interval between two consecutive gradient fields was decreased by a factor of 8 or 16 (depending on the number of scans used in NMR diffusion experiments).

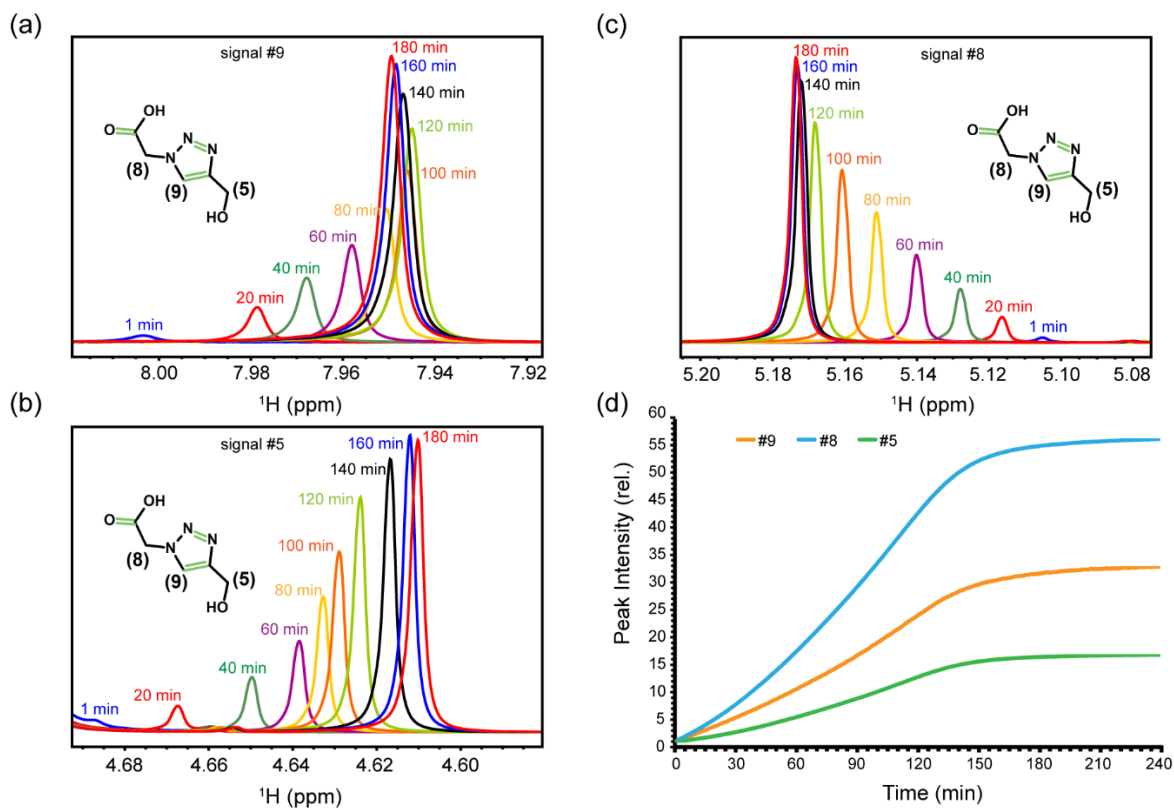

**Figure 3.** Kinetics of product formation during click reaction monitored through real-time 1D $^1$H NMR spectra. Time-dependent changes in three signals #5, #8 and #9 belonging to the product molecule (triazole) are shown in panels (a)-(c), along with its 2D chemical structure. In (a), note that the direction of chemical shift displacement is reversed after ca. 120 minutes of reaction. Time-dependent changes in NMR signal intensities are shown in (d).

First, we monitored how the effective diffusion coefficient ($D_{eff}$) of reactants **1** (alkyne) and **2** (azide) changed over the course of the CuAAC reaction. Both of the signals belonging to the alkyne molecule, i.e. signals #1 and #4, started with a $D_{eff}$ around 8% lower than the reference $D_0$ of 11.1*10$^{-10}$ m$^2$.s$^{-1}$ measured in the absence of reaction (Figure 4a). Along with the progress of the reaction, the two signals exhibited a further time-dependent drop in their $D_{eff}$, albeit with different patterns: the $D_{eff}$ associated to signal #1, i.e. the terminal alkyne proton, underwent three phases of rapid decay intervened by two phases of relative stability, while signal #4 exhibited an initial slow decrease in $D_{eff}$ followed by a more rapid drop. The time-dependent decay in mobility is in line with the formation of Cu-alkyne and 2Cu-alkyne complexes (species **2** and **3**, respectively in scheme 1b), which are significantly larger than the un-complexed alkyne molecule (species **1**), especially when we consider the (dynamic) network of ligand (e.g. water) molecules in the coordination sphere of copper ions.

Furthermore, as suggested by the reaction mechanism depicted in scheme 1b, the $D_{eff}$ of signal #1 represents the population-weighted average of the diffusion coefficients of species **1** and **2**, while the $D_{eff}$ of signal #4 is the corresponding average for species **1**, **2** and **3**. Consequently, it is not surprising that the two signals of the alkyne molecule followed distinct time-dependent changes in their $D_{eff}$ during the reaction. Similar to the signals of alkyne molecule, the azide signal #3 started with a $D_{eff}$ around 6% lower than the reference $D_0$ of 8.4*10$^{-10}$ m$^2$.s$^{-1}$. However, in line with the later entry of azide to the reaction cycle, its $D_{eff}$ remained nearly constant during the first 60-90 minutes of reaction (Figure 4b). It was then followed by clear time-dependent decay in mobility, as expected for reversible coordination of azide with 2Cu-alkyne complex (species **4**). The initial drop in $D_{eff}$ of azide is likely due to its known, albeit weak, coordination with copper ions.[36]

Next, the $D_{eff}$ of the product triazole molecule was probed through its well-resolved signals #8 and #9 (Figure 4c and SI, Figure S2). Interestingly, and in contrast with the two reactants, the triazole molecule exhibited a clear time-dependent increase in $D_{eff}$, so that the limiting value of $D_{eff}$ was ca. 9-10% larger than its starting value. The apparent diffusion enhancement towards the end of reaction can be explained considering the larger size of copper triazolide (species **6**) and especially copper metallacycle (species **5**) compared to the product triazole molecule.

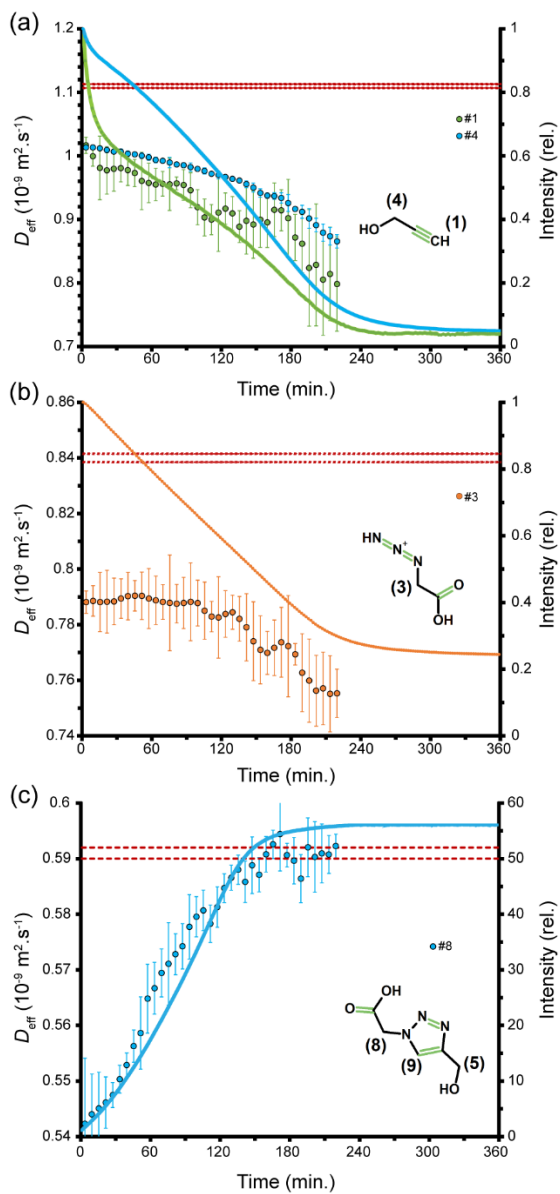

**Figure 4.** Diffusion of the reactants and product molecules during click reaction monitored through real-time PFG-NMR experiments. Both the alkyne (a) and azide (b) molecules begin with an effective diffusion coefficient ($D_{eff}$) smaller than their reference diffusion coefficients ($D_0$, average ± stdev, shown as dashed lines) and exhibit further decay during the reaction. The product molecule (c, triazole) however shows time-dependent rise in $D_{eff}$. The time-dependent changes in signal intensity are shown as lines.

Finally, we monitored the time dependence of the $D_{eff}$ of the catalyst ascorbate molecule using the well-resolved signal #2 (SI, Figure S3). Unlike the reactant alkyne and azide and product triazole molecules, the starting $D_{eff}$ of ascorbate molecule was slightly (ca. 2%) larger than the reference $D_0$ of $5.8*10^{-10}$ m$^2$.s$^{-1}$, however with the progress of reaction, its $D_{eff}$ slowly returned to the reference value. The initial increase in the mobility of ascorbate molecule is probably caused by the small increase in the temperature and its resultant decrease in sample viscosity.

To investigate whether the mobility alterations of alkyne are caused by copper π-coordination and/or σ-bind formation alone or further progression into reaction cycle underlies it, we studied a mixture of alkyne and catalyst (copper sulfate and sodium ascorbate), as in the original reaction mixture, but without azide. Consequent to the absence of azide, the reaction cycle would be stopped at step 2, where 2Cu-alkyne complex is formed. As shown in Figure 5, the signal #1 underwent gradual displacement towards upfield chemical shifts along with narrowing of the signal, which was later followed by intensity loss. Interestingly, the upfield direction of signal displacement was opposite to the downfield displacement observed during the full reaction (Figure 2a), indicating that the chemical species underlying signal #1 in the dissected and full reactions were different. Signal #4 however exhibited time-dependent chemical shift changes towards downfield similar to those observed in the full reaction. The $D_{eff}$ associated to signals #1 and #4 were ca. 3% lower than the reference $D_0$ of alkyne, however in contrast with the full reaction, they remained nearly constant along the reaction and did not show further drop. Taken together, our data point to the presence of multiple alkyne-copper species in rapid equilibrium with each other, and propose that an intermediate species other than copper alkyne complexes makes a significant contribution to alkyne mobility decay during the click reaction.

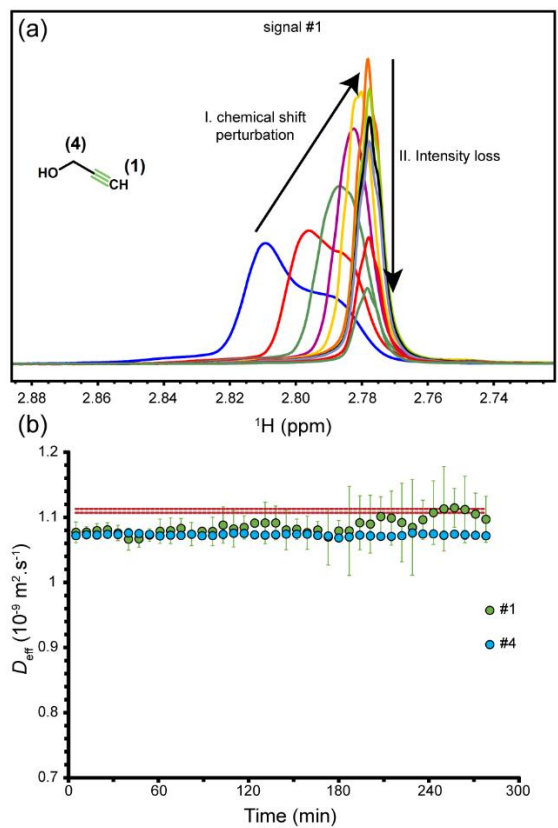

**Figure 5.** Cu(I)-coordination of alkyne in the absence of azide. In (a), time-dependent changes in the chemical shift and intensity of signal 1, belonging to terminal proton, are shown. As shown in (b), the effective diffusion coefficient of alkyne ($D_{eff}$) determined through its signals 1 and 4 is smaller than the reference $D_0$ value (average ± stdev, dashed line), and remains nearly constant during the coordination process.

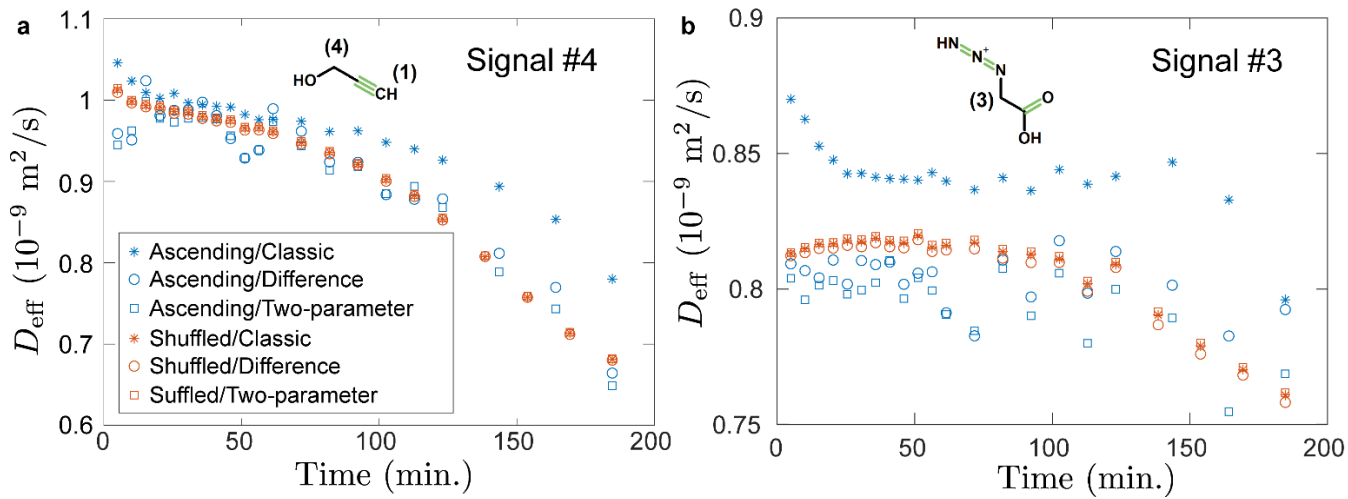

**Figure 6.** Comparison of PFG-NMR diffusion measurements performed using ascending (standard) and shuffled gradients, each of them analyzed using the "classic ST" equation, see eq. (S1), as well as the two modified ST equations ("difference" and "twoparameter") described in the main text, see eqs. (1) and (2). The signals correspond to the reactants alkyne (a) and azide (b). Notice how, when shuffled gradients are used (red symbols), all three ST equations give identical results. However, when ascending gradients are used (blue symbols), the classic ST equation overestimates the diffusion coefficient relative to the two modified ST equations, which in turn agree with each other and with the shuffled gradient data.

To verify the importance of using shuffled gradients in the presence of signal intensity changes, we also performed experiments in which gradients were ordered in increasing strength as in the standard NMR diffusion experiments. The diffusion coefficients obtained from a fit of the signal intensity *vs* gradient field strength data to the standard ST equation showed significant departures from those obtained using shuffled gradients. One may thus wonder whether it is possible to detect, and potentially correct for, the presence of the systematic artifacts due to signal intensity changes in an experiment that uses increasing gradient strengths. To this end, we devised two modified ST equations that account for the time-dependence of signal intensities *via* two independent methods, which we call "difference" and "two-parameter" method. In the difference method, we keep track of the order ($n = 1,2,3$ ...) in which the gradients $Q_n$ are applied, and consider only the difference between consecutive gradients, resulting in the modified ST equation

$$\frac{I_{n+1}}{I_n} = \exp[-D(Q_{n+1} - Q_n)] \quad \text{(eq. 1)}$$

which will not suffer from artifacts as long as that the signal intensity changes are negligible during the short interval between two consecutive gradient strengths, rather than during the whole measurement as the standard ST equation. In the two-parameter method, we use a first-order Taylor, i.e. linear, approximation of kinetics for the signal intensity changes during a full measurement starting at time $t$, so that $I_0(t + \Delta t) \approx I_0(t) + I_0'(t)\Delta t \equiv I_{0,t} + I_{0,t}'\Delta t$, and thus consider the modified ST equation

$$I(t + \Delta t, Q) = (I_{0,t} + I_{0,t}'\Delta t)\exp(-DQ) \quad \text{(eq. 2)}$$

which, besides the diffusion coefficient, gives information on the average rate $I_{0,t}'$ of signal intensity changes during the measurement. The linear approximation of the reaction kinetics is justified, considering the short duration of the NMR diffusion experiment compared to the timescale of reaction kinetics. Our results confirm that it is indeed possible to detect and correct for such artifacts (Figure 6). Whereas the standard ST and the two modified ST equations produced identical results when applied to the shuffled-gradient data, there was a substantial deviation between the results of the standard ST and the two modified ST equations for the increasing-gradient data. On the other hand, the two modified ST equations gave similar results to each other, and moreover brought the calculated diffusion coefficients closer to those obtained from the shuffled-gradients data.

**CONCLUSION**

Taken together, our results do not show evidence of boosted or active diffusion in the context of the click reaction. The changes in the measured diffusion coefficients over the course of the reaction for the NMR signals associated to reactants and products can be explained by the presence of relatively large intermediate species within the reaction cycle with lower diffusivities than both the reactants and the product molecules. The slight transient increase in diffusion observed for the catalyst can be explained as arising from changes in sample viscosity associated to a small temperature increase in the initial stages of the reaction. Moreover, we showed that is possible to detect and correct artifacts arising from signal intensity changes during a diffusion NMR experiment, even *a posteriori*, without the use of shuffled gradients.

The conclusions reached here for the molecular-scale click reaction mirror those that have been reached for a number of nano-scale catalytic enzymes, for which an initial claim of active enhanced diffusion was later shown to be a consequence of passive mechanisms such as conformational changes or subunit dissociation induced by substrate binding and/or catalysis, or even measurement artifacts.[16, 18, 19, 44, 45] From a theoretical perspective, it is important to bear in mind that the momentum imparted by any sudden impulse is dissipated almost instantaneously into the surrounding viscous medium and that any directed motion is randomized quickly by rotational Brownian motion.[15, 17-19]

Nevertheless, even in cases where the mechanism behind the diffusion changes is a passive one, unexpected effects can still arise in an out-of-equilibrium setting. For example, diffu-sion changes due to conformational changes or dissociation can cause directed motion and inhomogeneous steady states in the presence of gradients, and dissociating enzymes may reach and react faster with distant catalytic targets.[46, 47] In this regard, it is also worth noting that during a chemical reaction (even one maintained in a steady state) the populations of reaction intermediates, and therefore the effective diffusion coefficient of the participating molecules, are no longer bound to the constraints of thermodynamic equilibrium, and may be influenced by e.g. the relative kinetic rates of different reactions in the cycle, rather than by free energy differences only. Accounting for the different diffusivities of the various reaction intermediates will thus be important in any setting in which chemical reactions occur inhomogeneously in space, as is certainly the case in biological cells.

## ASSOCIATED CONTENT

**Supporting Information**. Description of Materials, Click Reaction and NMR experiments and Figures S1-S3.

## AUTHOR INFORMATION

### Corresponding Authors

* Nasrollah Rezaei-Ghaleh – Institut für Physikalische Biologie, Heinrich-Heine-Universität Düsseldorf, Universitätsstr. 1, D-40225 Düsseldorf, Germany;
Email: Nasrollah.Rezaie.Ghaleh@hhu.de
* Ramin Golestanian – Department of Living Matter Physics, Max Planck Institute for Dynamics and Self-Organization, Am Faßberg 11, D-37077 Göttingen, Germany;
Email: ramin.golestanian@ds.mpg.de

### Author Contributions

N.R.-G. and R.G. conceived of the project idea. N.R.-G. designed and conducted NMR experiments and analyzed data and wrote the manuscript. J.A.-C. developed the modified fitting methods and contributed to manuscript writing. C.G. and R.G. contributed to the development of ideas and commented on the manuscript. R.G. gave input regarding the overall structure of the manuscript and the interpretation of the results.

### Funding Sources

Deutsche Forschungsgemeinschaft (DFG, Research grants RE 3655/2-1 and RE 3655/2-3).

### Notes

The authors declare no competing financial interest.

## ACKNOWLEDGMENT

N.R.-G. acknowledges Deutsche Forschungsgemeinschaft (DFG, German Research Foundation) for research grants RE 3655/2-1 and RE 3655/2-3.

## ABBREVIATIONS

PFG-NMR, pulse field gradient-nuclear magnetic resonance
$D_{eff}$ and $D_0$, effective and reference diffusion coefficients
CuAAC, Cu(I)-catalyzed azide-alkyne cycloaddition
DFT, density functional theory.

## REFERENCES

1. Julicher, F.; Ajdari, A.; Prost, J., Modeling molecular motors. *Rev. Mod. Phys.* **1997**, *69*, 1269-1281.
2. Kay, E. R.; Leigh, D. A.; Zerbetto, F., Synthetic molecular motors and mechanical machines. *Angew. Chem. Int. Ed. Engl.* **2007**, *46*, 72-191.
3. Gompper, G.; Winkler, R. G.; Speck, T.; Solon, A.; Nardini, C.; Peruani, F.; Lowen, H.; Golestanian, R.; Kaupp, U. B.; Alvarez, L.; Kiorboe, T.; Lauga, E.; Poon, W. C. K.; DeSimone, A.; Muinos-Landin, S.; Fischer, A.; Soker, N. A.; Cichos, F.; Kapral, R.; Gaspard, P.; Ripoll, M.; Sagues, F.; Doostmohammadi, A.; Yeomans, J. M.; Aranson, I. S.; Bechinger, C.; Stark, H.; Hemelrijk, C. K.; Nedelec, F. J.; Sarkar, T.; Aryaksama, T.; Lacroix, M.; Duclos, G.; Yashunsky, V.; Silberzan, P.; Arroyo, M.; Kale, S., The 2020 motile active matter roadmap. *J. Phys.: Condens. Matter* **2020**, *32*, 193001.
4. Hortelao, A. C.; Simo, C.; Guix, M.; Guallar-Garrido, S.; Julian, E.; Vilela, D.; Rejc, L.; Ramos-Cabrer, P.; Cossio, U.; Gomez-Vallejo, V.; Patino, T.; Llop, J.; Sanchez, S., Swarming behavior and in vivo monitoring of enzymatic nanomotors within the bladder. *Sci. Robot.* **2021**, *6*, eabd2823.
5. de Avila, B. E. F.; Angsantikul, P.; Li, J. X.; Lopez-Ramirez, M. A.; Ramirez-Herrera, D. E.; Thamphiwatana, S.; Chen, C. R.; Delezuk, J.; Samakapiruk, R.; Ramez, V.; Obonyo, M.; Zhang, L. F.; Wang, J., Micromotor-enabled active drug delivery for in vivo treatment of stomach infection. *Nat. Commun.* **2017**, *8*, 272.
6. Golestanian, R.; Liverpool, T. B.; Ajdari, A., Designing phoretic micro- and nano-swimmers. *New J. Phys.* **2007**, *9*, 126.
7. Golestanian, R., Phoretic Active Matter. **2019**, arXiv:1909.03747 [cond-mat.soft].
8. Howse, J. R.; Jones, R. A. L.; Ryan, A. J.; Gough, T.; Vafabakhsh, R.; Golestanian, R., Self-motile colloidal particles: From directed propulsion to random walk. *Phys. Rev. Lett.* **2007**, *99*, 048102.
9. Golestanian, R., Anomalous Diffusion of Symmetric and Asymmetric Active Colloids. *Phys. Rev. Lett.* **2009**, *102*, 188305.
10. Golestanian, R., Synthetic Mechanochemical Molecular Swimmer. *Phys. Rev. Lett.* **2010**, *105*, 018103.
11. Golestanian, R.; Ajdari, A., Mechanical response of a small swimmer driven by conformational transitions. *Phys. Rev. Lett.* **2008**, *100*, 038101.
12. Jee, A. Y.; Tlusty, T.; Granick, S., Master curve of boosted diffusion for 10 catalytic enzymes. *Proc. Natl. Acad. Sci. USA* **2020**, *117*, 29435-29441.
13. Muddana, H. S.; Sengupta, S.; Mallouk, T. E.; Sen, A.; Butler, P. J., Substrate Catalysis Enhances Single-Enzyme Diffusion. *J. Am. Chem. Soc.* **2010**, *132*, 2110-2111.
14. Riedel, C.; Wilson, C. W. A.; Hamadani, K.; Tsekouras, K.; Marqusee, S.; Presse, S.; Bustamante, C., The Heat Released by a Chemical Reaction Locally Enhanced the Enzyme Diffusion. *Biophys. J.* **2014**, *106*, 668a-668a.
15. Golestanian, R., Enhanced Diffusion of Enzymes that Catalyze Exothermic Reactions. *Phys. Rev. Lett.* **2015**, *115*, 108102.
16. Agudo-Canalejo, J.; Adeleke-Larodo, T.; Illien, P.; Golestanian, R., Enhanced Diffusion and Chemotaxis at the Nanoscale. *Acc. Chem. Res.* **2018**, *51*, 2365-2372.
17. Feng, M. D.; Gilson, M. K., A Thermodynamic Limit on the Role of Self-Propulsion in Enhanced Enzyme Diffusion. *Biophys. J.* **2019**, *116*, 1898-1906.
18. Gunther, J. P.; Borsch, M.; Fischer, P., Diffusion Measurements of Swimming Enzymes with Fluorescence Correlation Spectroscopy. *Acc. Chem. Res.* **2018**, *51*, 1911-1920.
19. Zhang, Y. F.; Hess, H., Enhanced Diffusion of Catalytically Active Enzymes. *ACS Cent. Sci.* **2019**, *5*, 939-948.
20. Pavlick, R. A.; Dey, K. K.; Sirjoosingh, A.; Benesi, A.; Sen, A., A catalytically driven organometallic molecular motor. *Nanoscale* **2013**, *5*, 1301-1304.
21. Dey, K. K.; Pong, F. Y.; Breffke, J.; Pavlick, R.; Hatzakis, E.; Pacheco, C.; Sen, A., Dynamic Coupling at the Angstrom Scale. *Angew. Chem. Int. Ed. Engl.* **2016**, *55*, 1113-1117.
22. MacDonald, T. S. C.; Price, W. S.; Astumian, R. D.; Beves, J. E., Enhanced Diffusion of Molecular Catalysts is Due to Convection. *Angew. Chem. Int. Ed. Engl.* **2019**, *58*, 18864-18867.
23. Wang, H.; Park, M.; Dong, R.; Kim, J.; Cho, Y. K.; Tlusty, T.; Granick, S., Boosted molecular mobility during common chemical reactions. *Science* **2020**, *369*, 537-541.

24. Zhang, Y. F.; Hess, H., Chemically-powered swimming and diffusion in the microscopic world. *Nat. Rev. Chem.* **2021,** *5*, 500-510.

25. Gunther, J. P.; Fillbrook, L. L.; MacDonald, T. S. C.; Majer, G.; Price, W. S.; Fischer, P.; Beves, J. E., Comment on "Boosted molecular mobility during common chemical reactions". *Science* **2021,** *371*, eabe8322.

26. Wang, H.; Park, M.; Dong, R.; Kim, J.; Cho, Y. K.; Tlusty, T.; Granick, S., Response to Comment on "Boosted molecular mobility during common chemical reactions". *Science* **2021,** *371*, eabe8678.

27. Wang, H.; Huang, T.; Granick, S., Using NMR to Test Molecular Mobility during a Chemical Reaction. *J. Phys. Chem. Lett.* **2021,** *12*, 2370-2375.

28. Fillbrook, L. L.; Gunther, J. P.; Majer, G.; Price, W. S.; Fischer, P.; Beves, J. E., Comment on "Using NMR to Test Molecular Mobility during a Chemical Reaction". *J. Phys. Chem. Lett.* **2021,** *12*, 5932-5937.

29. Huang, T.; Wang, H.; Granick, S., Reply to Comment on "Using NMR to Test Molecular Mobility during a Chemical Reaction". *J. Phys. Chem. Lett.* **2021,** *12*, 5744-5747.

30. Huang, T.; Li, B.; Wang, H.; Granick, S., Molecules, the Ultimate Nanomotor: Linking Chemical Reaction Intermediates to their Molecular Diffusivity. *ACS Nano* **2021,** *15*, 14947-14953.

31. Fillbrook, L. L.; Günther, J. P.; Majer, G.; O'Leary, D.; Price, W. S.; Fischer, P.; Beves, J. E., Following Molecular Mobility During Chemical Reactions: No Evidence for Active Propulsion. ChemRxiv. Cambridge: Cambridge Open Engage; **2021**; This content is a preprint and has not been peer-reviewed.

32. Johnson, C. S., Diffusion ordered nuclear magnetic resonance spectroscopy: principles and applications. *Prog. Nucl. Magn. Reseon. Spectrsoc.* **1999,** *34*, 203-256.

33. MacDonald, T. S. C.; Price, W. S.; Beves, J. E., Time-Resolved Diffusion NMR Measurements for Transient Processes. *ChemPhysChem* **2019,** *20*, 926-930.

34. Tornoe, C. W.; Christensen, C.; Meldal, M., Peptidotriazoles on solid phase: [1,2,3]-triazoles by regiospecific copper(I)-catalyzed 1,3-dipolar cycloadditions of terminal alkynes to azides. *J. Org. Chem.* **2002,** *67*, 3057-3064.

35. Rostovtsev, V. V.; Green, L.; Sharpless, K. B., Copper-catalyzed cycloaddition of azides and acetylenes. *Abstr. Pap. Am. Chem. Soc.* **2002,** *224*, U186-U186.

36. Hein, J. E.; Fokin, V. V., Copper-catalyzed azide-alkyne cycloaddition (CuAAC) and beyond: new reactivity of copper(I) acetylides. *Chem. Soc. Rev.* **2010,** *39*, 1302-1315.

37. Himo, F.; Lovell, T.; Hilgraf, R.; Rostovtsev, V. V.; Noodleman, L.; Sharpless, K. B.; Fokin, V. V., Copper(I)-catalyzed synthesis of azoles. DFT study predicts unprecedented reactivity and intermediates. *J. Am. Chem. Soc.* **2005,** *127*, 210-216.

38. Worrell, B. T.; Malik, J. A.; Fokin, V. V., Direct Evidence of a Dinuclear Copper Intermediate in Cu(I)-Catalyzed Azide-Alkyne Cycloadditions. *Science* **2013,** *340*, 457-460.

39. Ahlquist, M.; Fokin, V. V., Enhanced reactivity of dinuclear Copper(I) acetylides in dipolar cycloadditions. *Organometallics* **2007,** *26*, 4389-4391.

40. Nolte, C.; Mayer, P.; Straub, B. F., Isolation of a copper(I) triazolide: A"click" intermediate". *Angew. Chem. Int. Ed. Engl.* **2007,** *46*, 2101-2103.

41. Straub, B. F., mu-acetylide and mu-alkenylidene ligands in "click" triazole syntheses. *Chem. Commun.* **2007,** (37), 3868-3870.

42. Swan, I.; Reid, M.; Howe, P. W. A.; Connell, M. A.; Nilsson, M.; Moore, M. A.; Morris, G. A., Sample convection in liquid-state NMR: Why it is always with us, and what we can do about it. *J. Magn. Reson.* **2015,** *252*, 120-129.

43. Webb, A. G., In *Annual Reports on NMR Spectroscopy*, Academic Press: 2002; Vol. 45, pp 1-67.

44. Jee, A. Y.; Chen, K.; Tlusty, T.; Zhao, J.; Granick, S., Enhanced Diffusion and Oligomeric Enzyme Dissociation. *J. Am. Chem. Soc.* **2019,** *141*, 20062-20068.

45. Chen, Z. J.; Shaw, A.; Wilson, H.; Woringer, M.; Darzacq, X.; Marqusee, S.; Wang, Q.; Bustamante, C., Single-molecule diffusometry reveals no catalysis-induced diffusion enhancement of alkaline phosphatase as proposed by FCS experiments. *Proc. Natl. Acad. Sci. USA* **2020,** *117*, 21328-21335.

46. Agudo-Canalejo, J.; Illien, P.; Golestanian, R., Phoresis and Enhanced Diffusion Compete in Enzyme Chemotaxis. *Nano Lett.* **2018,** *18*, 2711-2717.

47. Agudo-Canalejo, J.; Illien, P.; Golestanian, R., Cooperatively enhanced reactivity and "stabilitaxis" of dissociating oligomeric proteins. *Proc. Natl. Acad. Sci. USA* **2020,** *117*, 11894-11900.

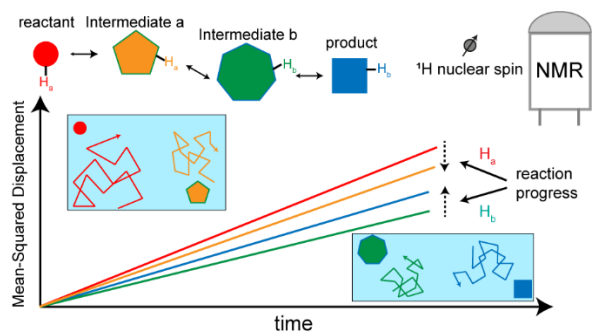

## Supporting Information

### Materials
Propyl-2-ny-ol (propargyl alcohol), sodium ascorbate, copper(II) sulfate pentahydrate and trimethylsilane (TMS) were from Sigma-Aldrich and 2-azido-aceticacid from Jena Bioscience. Deuterated solvent (>99.95% $D_2O$) was from Cambridge Isotope Laboratories.

### CuCAA click reaction
The Cu(I)-catalyzed Alkyne-Azide Cycloaddition (CuAAC) click reaction was triggered by adding 16 mM copper sulfate and 64 mM sodium ascorbate to a mixture of 200 mM propyl-2-ny-ol (henceforth, alkyne) and 200 mM 2-Azido acetic acid in $D_2O$.[38] The final volume of reaction mixture was 160 μL. The reaction mixture was then immediately transferred to NMR tubes for measurements. The stock solution of sodium ascorbate was freshly prepared each time, in order to minimize its oxidative degradation.

### NMR experiments
NMR measurements were performed on an 800 MHz Bruker (Germany) spectrometer, equipped with a cryogenic TCI probe with a z-axis gradient coil. The standard NMR tubes (hilgenberg, Germany) with inner (ID) and outer diameters (OD) of respectively 2.36 and 2.95 mm were used for NMR measurements. The NMR experiments were performed at 298 K, for which the temperature was controlled to ±0.05 K using the Bruker VT unit calibrated through a standardized thermocouple. The NMR samples contained 100% $D_2O$, in which the deuteron NMR signal was used for frequency locking. Whenever needed, a small amount of TMS was added into NMR sample and its proton signal was used for chemical shift referencing (0.000 ppm).

The kinetics of click reaction were monitored through standard real-time 1D $^1H$ pulse-acquire experiments, in which the first NMR experiment was started shortly (3-5 minutes) after addition of catalysts, i.e. copper sulfate and ascorbate sodium, to reaction mixture. To ensure an almost complete recovery of longitudinal magnetization for different protons (of reactants, catalysts, intermediates and product), a relatively long recycle delay of 10 s was used in kinetic experiments. The average temperature of NMR samples during click reaction was determined using the known temperature-dependence of the chemical shift of residual HDO proton signal (referenced with TMS proton signal, see above).[43]

Pulse field gradient (PFG)-NMR diffusion experiments were measured using the standard or modified "stimulated echo bipolar gradient pulse pairs with one coil" (stebpgp1s) or the convection-compensating "double-stimulated echo bipolar gradient pulse pairs with three coils" (dstebpgp3s) sequences.[42] The diffusion delay (big delta, $\Delta$) of 25 ms and diffusion gradient length (little delta, $\delta$) of 2.25 ms were used for diffusion experiments. These values were chosen after exploring the parameter space (big and little deltas ranging over 25-100 ms and 2.25-3.5 ms, respectively) and showing that at the chosen diffusion times the diffusion coefficients obtained through standard and convection-compensating sequences were effectively identical. To account for the time-dependent changes in NMR signal intensities due to click reaction (reactant consumption, product formation) during PFG-NMR experiment, the stebpgp1s and dstebpgp3s sequences were modified to allow the application of gradient field strengths in a random "shuffled" order, and shorten the time interval between different gradient strengths through single-scan-interleaved acquisition scheme. For the sake of comparison, the PFG-NMR experiments were also conducted in the standard manner, i.e. with gradients ramping linearly from 5 to 95%. The number of scans were 8 and 16, respectively for stebpgp1s and convection-compensating dstebpgp3s NMR pulse sequences, with a total recycle delay ($acq+d_1$) of 3 s and duration of each PFG-NMR experiment of 6 and 10 minutes.

To obtain the diffusion coefficients ($D$), the gradient-dependent intensity attenuation of NMR signals were fitted to Stejskal-Tanner (ST) equation,

$$I = I_0 \exp(-DQ) \quad \text{(eq. S1)}$$

with $Q = \gamma^2 \delta^2 \left(\Delta - \frac{\delta}{3}\right) g^2$, where $\gamma$ is gyromagnetic ratio of proton, $g$ is gradient strength and little and big delta ($\delta$, $\Delta$) are as defined above. The fitted data are presented as three-point averages, where each point is weighted (inversely) by the sum of squared errors (SSE) of the fits. The magnetic field gradient was calibrated using a 99.95% $D_2O$ sample and the known diffusion coefficient of HDO molecules at 25 °C ($1.900 \pm 0.004 \times 10^{-10}$ $m^2.s^{-1}$).[48] To account for the time-dependence of reactants and product signal intensities during the catalyzed reaction, the NMR diffusion data were also analyzed using two modified versions of ST equations (see the main text).

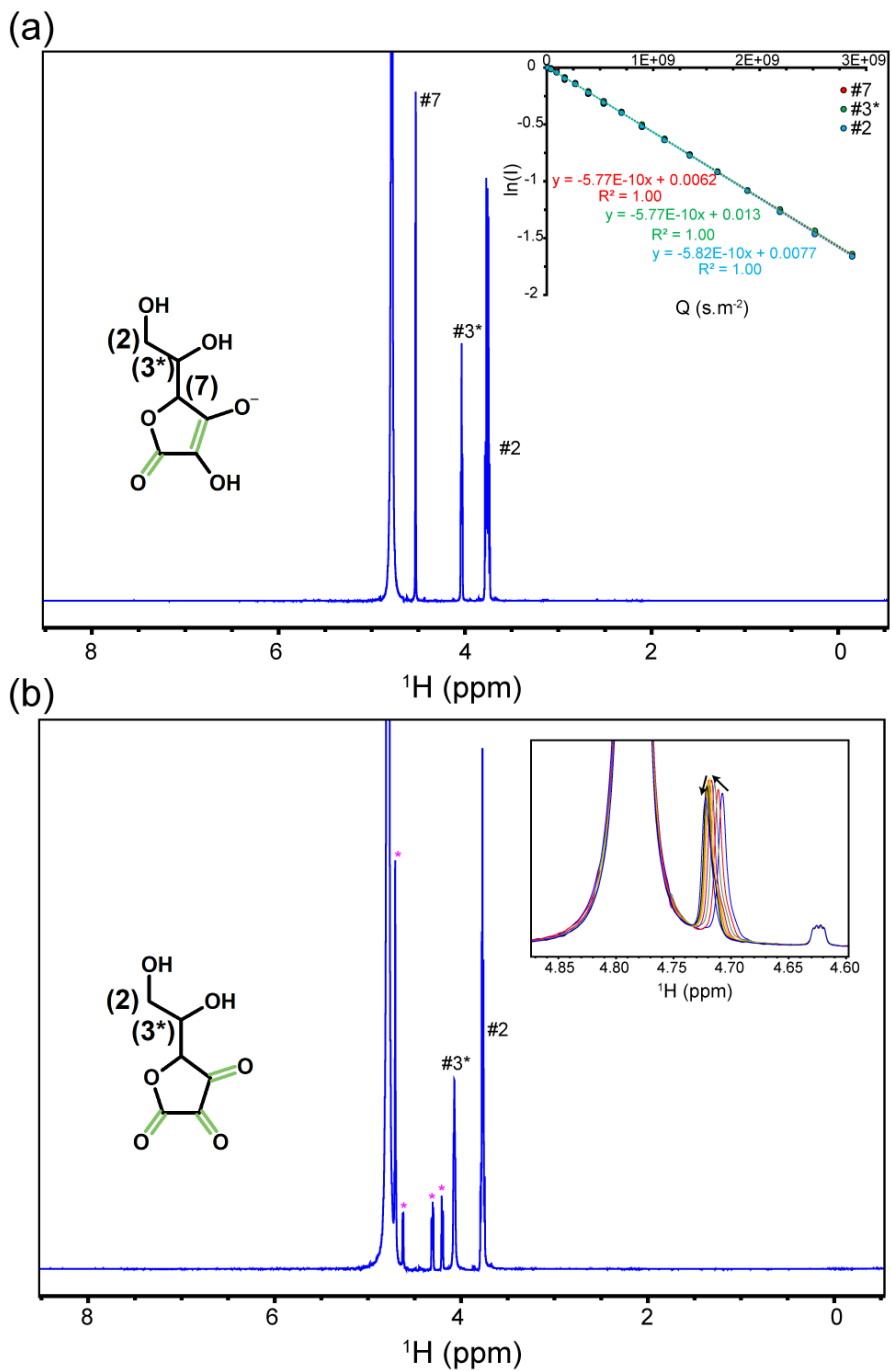

**Figure S1.** 1D $^1$H NMR spectra of the catalyst of click reaction (sodium ascorbate, 64 mM), measured in the absence (a) or presence of 16 mM CuSO4 (b). Upon addition of CuSO$_4$, the signal 7 disappears and four new signals, probably belonging to the oxidation products such as dehydroascorbic acid, emerge (marked with pink asterisks). The NMR signals are assigned according to the shown 2D chemical structures of ascorbate and dehydroascorbic acid. In (a), the Inset shows the gradient-dependent NMR intensity attenuation in log-quadratic scale for three signals of ascorbate molecule, where the linear slopes represent the reference diffusion coefficient ($D_0$) of ascorbate molecule. The Inset of (b) shows time-dependent changes in one of the newly emerged peaks following CuSO$_4$ addition.

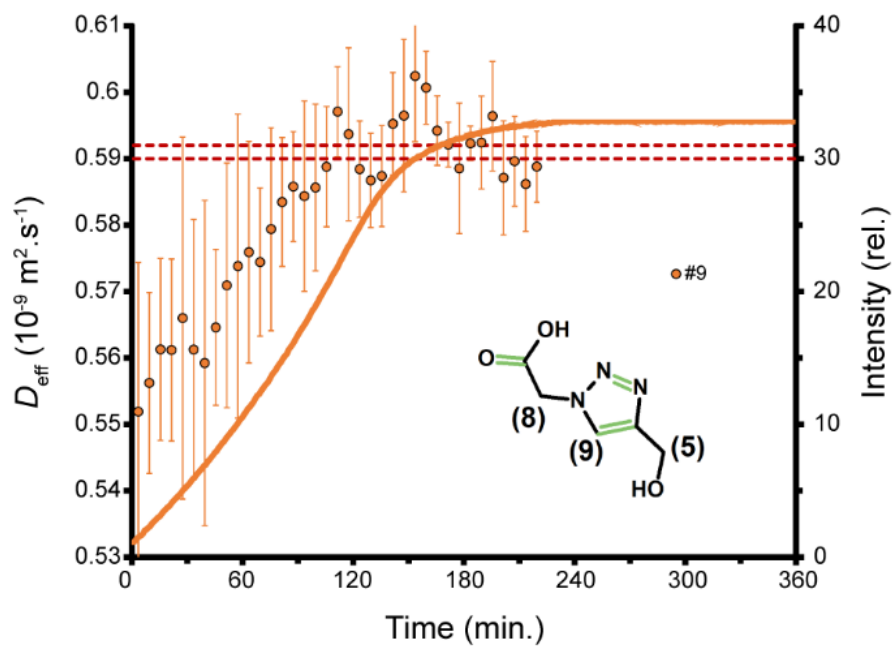

**Figure S2.** Diffusion of the product molecule during click reaction monitored through real-time PFG-NMR experiments. The effective diffusion coefficient ($D_{eff}$) of product molecule (triazole) determined via its NMR signal 9 exhibits gradual rise over the course of click reaction (see also Fig. 4c). The reference diffusion coefficient ($D_0$, average ± stdev) of triazole molecule is shown as dashed lines. The kinetics of click reaction is represented by changes in the intensity of signal 9 (solid line).

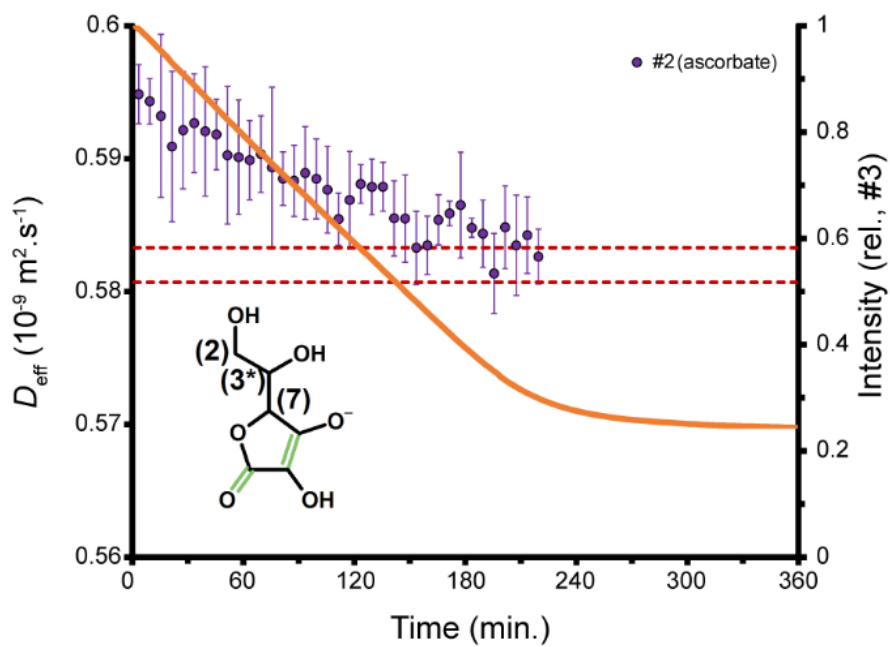

**Figure S3.** Diffusion of the catalyst ascorbate molecule during click reaction monitored through real-time PFG-NMR experiments. In the beginning of click reaction, the effective diffusion coefficient ($D_{eff}$) of ascorbate is slightly (ca. 2%) larger than the reference diffusion coefficient ($D_0$, average ± stdev shown as dashed lines), but slowly decays towards the reference value along with the progression of click reaction. The kinetics of click reaction is represented by changes in the intensity of signal 3 belonging to the reactant azide molecule (solid line).

## Supporting References

1.    Worrell, B. T.; Malik, J. A.; Fokin, V. V., Direct Evidence of a Dinuclear Copper Intermediate in Cu(I)-Catalyzed Azide-Alkyne Cycloadditions. *Science* **2013,** *340*, 457-460.
2.    Webb, A. G., In *Annual Reports on NMR Spectroscopy*, Academic Press: 2002; Vol. 45, pp 1-67.
3.    Swan, I.; Reid, M.; Howe, P. W. A.; Connell, M. A.; Nilsson, M.; Moore, M. A.; Morris, G. A., Sample convection in liquid-state NMR: Why it is always with us, and what we can do about it. *J. Magn. Reson.* **2015,** *252*, 120-129.
4.    Mills, R., Self-Diffusion in Normal and Heavy-Water in Range 1-45 Degrees. *J. Phys. Chem.-Us* **1973,** *77*, 685-688.